\documentclass[pra,twocolumn,floatfix,superscriptaddress,longbibliography,notitlepage]{revtex4-2}
\usepackage[T1]{fontenc}
\usepackage[utf8]{inputenc}
\usepackage{amssymb,amsmath,amsthm,color,times,graphicx}
\usepackage{caption}
\usepackage{ragged2e}
\DeclareCaptionJustification{justified}{\justifying}
\usepackage{graphicx}
\newcommand{\PT}{\mathcal{P}\mathcal{T}}
\usepackage{xcolor}
\usepackage{tikz}
\usepackage{booktabs}
\usepackage{graphicx}
\usepackage{multirow}
\usepackage{bbm}
\usepackage[colorlinks=true,%
            linkcolor=blue,    
            citecolor=blue,    
            urlcolor=blue]{hyperref}   
            
\definecolor{norange}{RGB}{230,120,20}

\usepackage{braket}
\usepackage{subcaption}
\usepackage{cleveref}

\begin{document}
\title{Quantum-inspired dynamical models on quantum and classical annealers}

\author{Philipp Hanussek}
\email{philipp.hanussek@etu.sorbonne-universite.fr}
\affiliation{Institute of Theoretical and Applied Informatics, Polish Academy of Sciences, Ba\l{t}ycka 5, Gliwice, 44-100, Poland}
\affiliation{Sorbonne Université, Faculté des Sciences et Ingénierie, 75005 Paris, France}
\author{Jakub Paw\l{o}wski}
\email{jakub.pawlowski@pwr.edu.pl}
\affiliation{Institute of Theoretical Physics, Faculty of Fundamental Problems of Technology, Wroc\l{a}w University of Science and Technology, 50-370, Wroc\l{a}w, Poland}
\affiliation{Quantumz.io Sp. z o.o., Pu\l{a}wska 12/3, 02-566, Warsaw, Poland}
\author{Zakaria Mzaouali}
\email{z.mzaouali@extern.fz-juelich.de}
\affiliation{J\"ulich Supercomputing Centre, Institute for Advanced Simulation, Forschungszentrum J\"ulich, Wilhelm-Johnen-Stra{\ss}e, J\"ulich, 52428, Germany.}
\affiliation{Institut für Theoretische Physik, Universität Tübingen, Auf der Morgenstelle 14, 72076 Tübingen, Germany}
\author{Bart\l{o}miej Gardas}
\email{bgardas@iitis.pl}
\affiliation{Institute of Theoretical and Applied Informatics, Polish Academy of Sciences, Ba\l{t}ycka 5, Gliwice, 44-100, Poland}

\date{\today}

\begin{abstract}
We propose a practical, physics-inspired benchmarking suite to challenge both quantum and classical computers by mapping real-time quantum dynamics to a common optimization format. Using a parallel-in-time encoding, we convert the real-time propagator of an $n$-qubit, possibly non-Hermitian, Hamiltonian into quadratic unconstrained binary optimization (QUBO) instances that are executable in a solver-agnostic manner on quantum annealers and classical optimizers alike. This enables direct, like-for-like performance comparisons across fundamentally different computational paradigms.
To stress-test the framework, we consider eight representative dynamical models spanning single-qubit rotations, multi-qubit entangling gates (Bell, GHZ, cluster), and PT-symmetric and other non-Hermitian generators, and evaluate success probability and time-to-solution as standard benchmarking metrics. Applying this methodology to two generations of D-Wave quantum annealers and to state-of-the-art classical solvers (Simulated Annealing and the GPU-accelerated VeloxQ), we find that Advantage2 consistently outperforms its predecessor, while VeloxQ retains the shortest absolute runtimes, reflecting the maturity of classical heuristics.
We further extend the benchmarks to large-scale instances ($N \simeq 10^{5}$), establishing a demanding classical baseline for future hardware. Together, these results position the parallel-in-time QUBO framework as a versatile and physically motivated testbed for quantitatively tracking progress toward quantum-competitive simulation of dynamical systems.
\end{abstract}

\maketitle
\section{Introduction}
Simulating the real-time evolution of many-body quantum systems is central to condensed matter physics~\cite{lanyon2011, bernien2017, andersen2025}, quantum chemistry~\cite{aspuru2005, peruzzo2014, georges2025}, and emerging quantum technologies~\cite{euchner2025}. However, the exponential growth of the Hilbert space, makes classical simulations infeasible beyond a few dozen qubits. This limitation motivated the idea that quantum hardware could outperform classical methods in simulating quantum dynamics \cite{Feynman1999, Lloyd1996, lanes2025}. Quantum annealers, such as those developed by D-Wave Systems, offer a platform for exploring such advantages. Recent experiments have suggested that they can simulate quantum dynamics beyond capabilities of classical devices~\cite{king2025, lidar_advantage}, albeit these results are still a subject of debate~\cite{Sels2025, Carleo2025, response_to_lidar}.

\begin{figure*}
    
    \includegraphics[width=\linewidth]{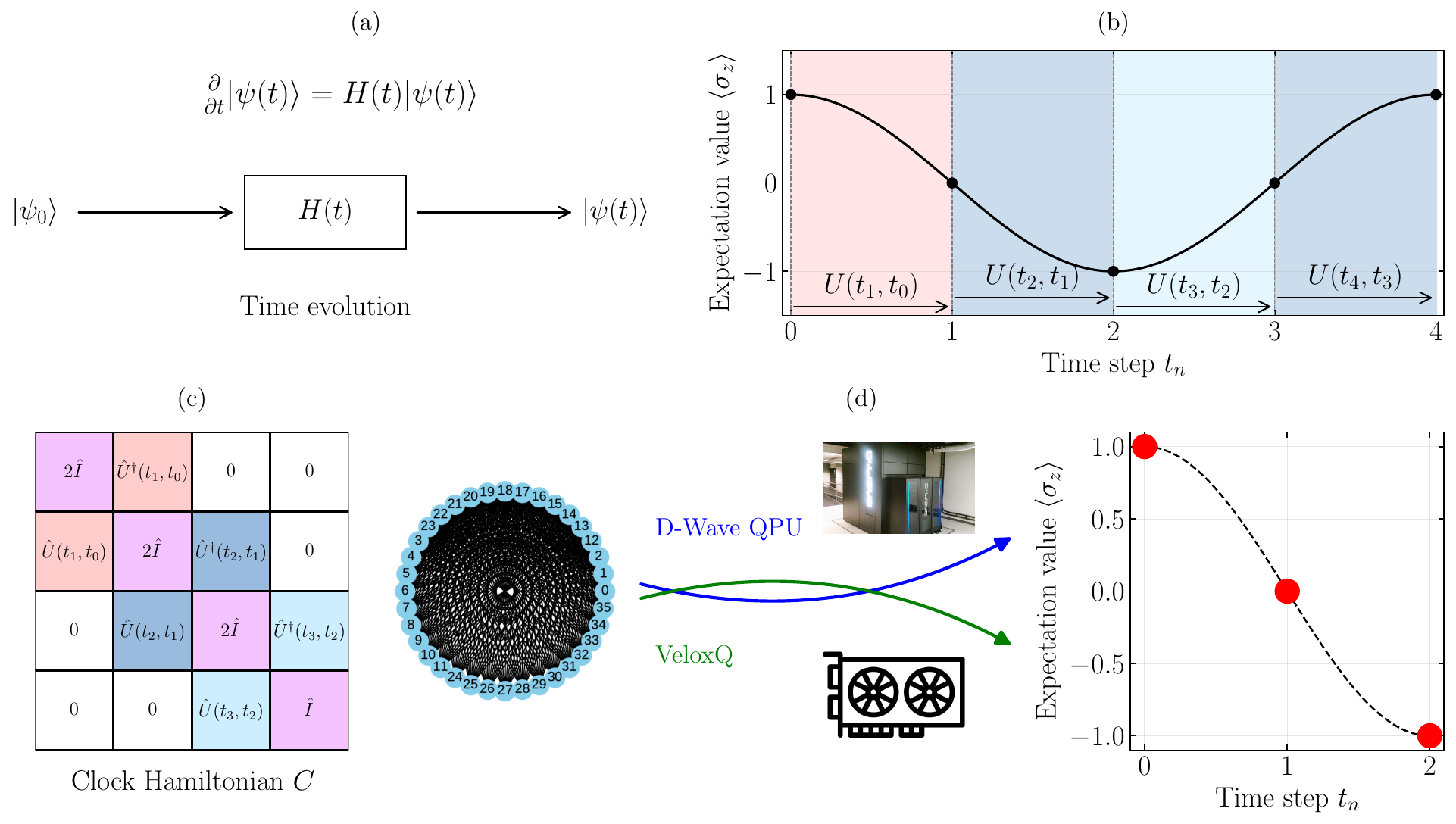}
    \caption{\textbf{Overview of the paper workflow.} (a) A quantum state \(\ket{\psi_0}\) of dimension \(N\) evolves under the (possibly non-Hermitian) generator \(H(t)\) according to the Schrödinger equation
\(\partial_t\ket{\psi(t)} = H(t)\ket{\psi(t)}\). (b) The real-time interval \([t_0,t_f]\) is discretised into \(N\) equal Trotter slices of width \(\Delta t\); each slice is propagated by the short-time operator
\(U(t_{n+1},t_n)=\exp\bigl[H(t_n)\,\Delta t\bigr]\), so that the sequence \(\{U(t_{n+1},t_n)\}_{n=0}^{N-2}\) fully specifies the history. (c) Discretised dynamics are encoded in the block-tridiagonal clock Hamiltonian \(C\), Eq.~\eqref{clock}; appending the initial-state projector yields the positive-definite matrix \(A\). Writing the quadratic form \(\tfrac12\langle x|A|x\rangle-\langle x\ket{\phi}\) in fixed-point binary variables (\(R\) bits over the range \([-2^{D},2^{D}]\)) produces a quadratic-unconstrained binary optimisation
(QUBO) with \(\mathcal{O}(LNR)\) spins, Eq.~\eqref{qubo}.
(d) The QUBO instance is submitted to a D-Wave quantum annealer (Advantage and Advantage2) and to the GPU-accelerated classical heuristic \textsc{VeloxQ}~\cite{veloxq}.
Both solvers return low-energy spin configurations that decode unambiguously into the time-ordered amplitudes \(\{|\psi(t_n)\rangle\}_{n=0}^{N-1}\);
from these, physical observables---e.g.\ \(\langle\sigma_z\rangle(t)\) shown on the right---are reconstructed without further fitting. Both solvers produce the same observable dynamics. This pipeline therefore maps continuous-time quantum dynamics to a solver-agnostic combinatorial problem, enabling a like-for-like performance comparison between quantum and classical optimisation hardware.}
    \label{fig:overview}
\end{figure*}
In the current noisy intermediate-scale quantum (NISQ) era, decisive demonstrations of advantage have proved elusive~\cite{arute2019,wu2021,kim2023,zhong2020,madsen2022,hangleiter2022}. Gate-based processors suffer from limited qubit counts and decoherence, while analogue platforms, such as ultracold atoms, trapped ions and photonic devices, excel at modelling specific Hamiltonians but lack flexibility~\cite{Michielsen_2017, fauseweh2024quantum, Georgescu2014}. Quantum annealers, produced most prominently by D-Wave Systems, occupy a specific niche. Their main task is the minimisation of a quadratic-unconstrained binary optimisation (QUBO) cost functions on fabricated Ising graphs, offering thousands of physical qubits that operate continuously rather than in discrete gate cycles. Because the hardware is already available as a cloud service, annealers are attractive testbeds for near-term advantage claims~\cite{king2025}.

Yet two technical hurdles have so far blocked a rigorous assessment of annealers for dynamical simulation~\cite{Vrinda2025}. First, mapping Schrödinger's equation-fundamentally a sequence of unitary operators-onto a static QUBO cost function has lacked an efficient, scalable recipe. Second, without agreed-upon benchmark suites it is impossible to separate genuine quantum speed-ups from artefacts caused by, e.g., a particularly easy instance or a fortuitous embedding on the hardware graph. Consequently, previous performance studies have either relied on synthetic optimisation problems that are only indirectly related to real-time quantum mechanics or compared quantum and classical solvers on disjoint workloads.


Our work contributes in closing this gap via a rigorous benchmarking protocol built upon a parallel in time encoding of quantum dynamics. The same family of instances is executed (i) on two generations of D-Wave quantum annealers-Advantage (Pegasus topology) and the newer Advantage2 (Zephyr topology with 33\% higher connectivity), (ii) on VeloxQ, a GPU-accelerated classical heuristic that solves the identical QUBOs but without hardware-embedding constraints~\cite{veloxq} and (iii) using both CPU and GPU implenentations of simulated annealing, which serves as the well known point of reference for classical solvers. This like-for-like design removes confounding factors and exposes the scaling behaviour that ultimately determines whether-and when-quantum advantage can
emerge.
\begin{figure*}
\subfloat[\label{anneal_schedule}]{\includegraphics[width=0.33\linewidth]{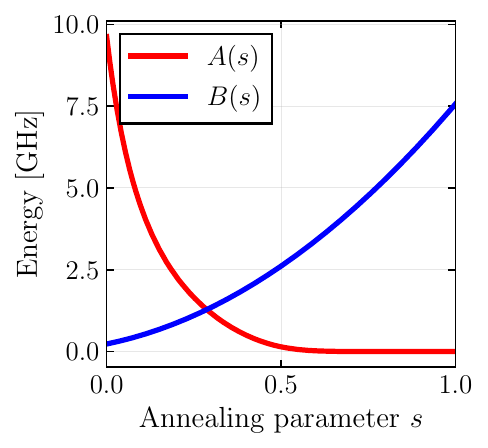}}%
    \subfloat[\label{pegasus}]{\includegraphics[width=0.33\linewidth]{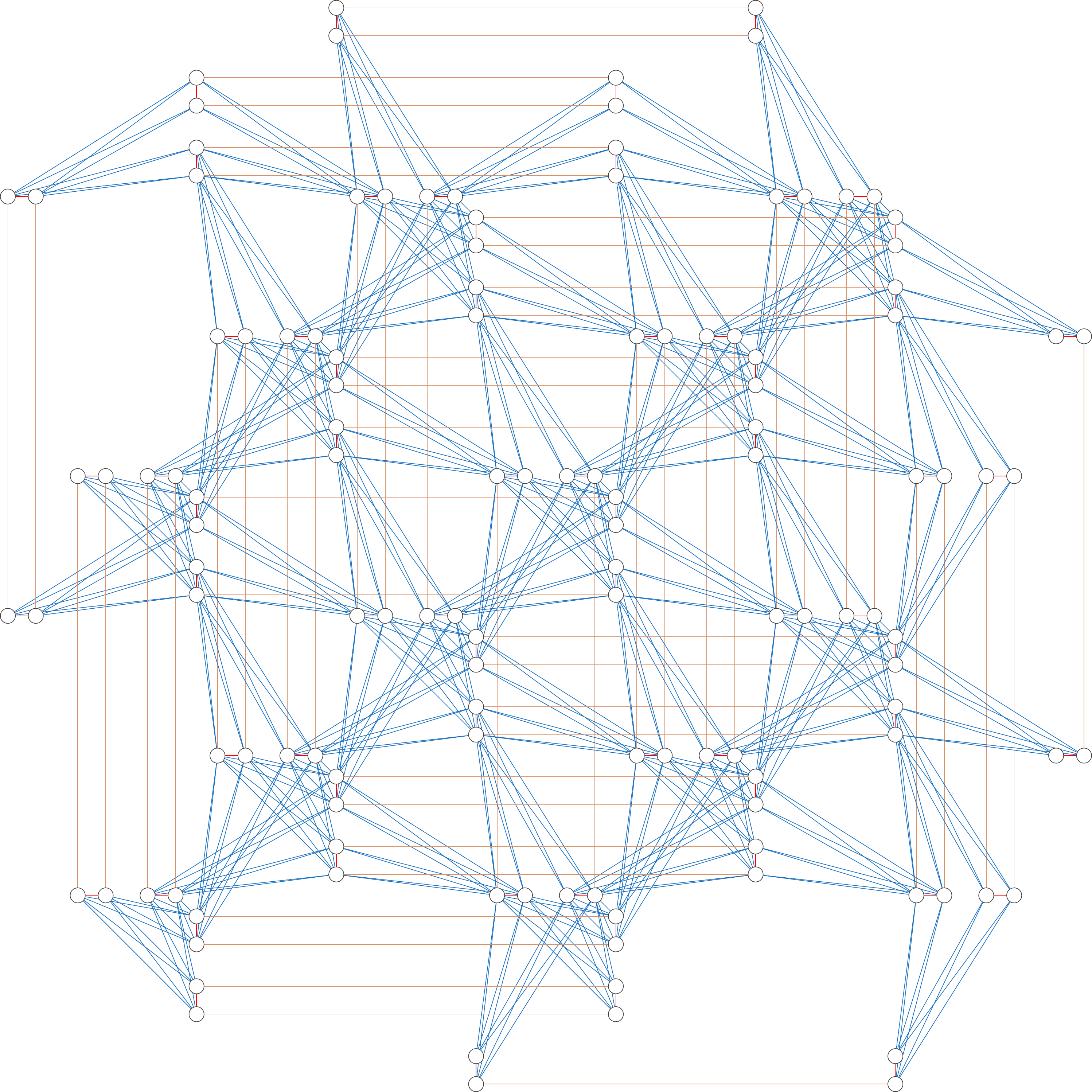}}%
    \subfloat[\label{zephyr}]{\includegraphics[width=0.33\linewidth]{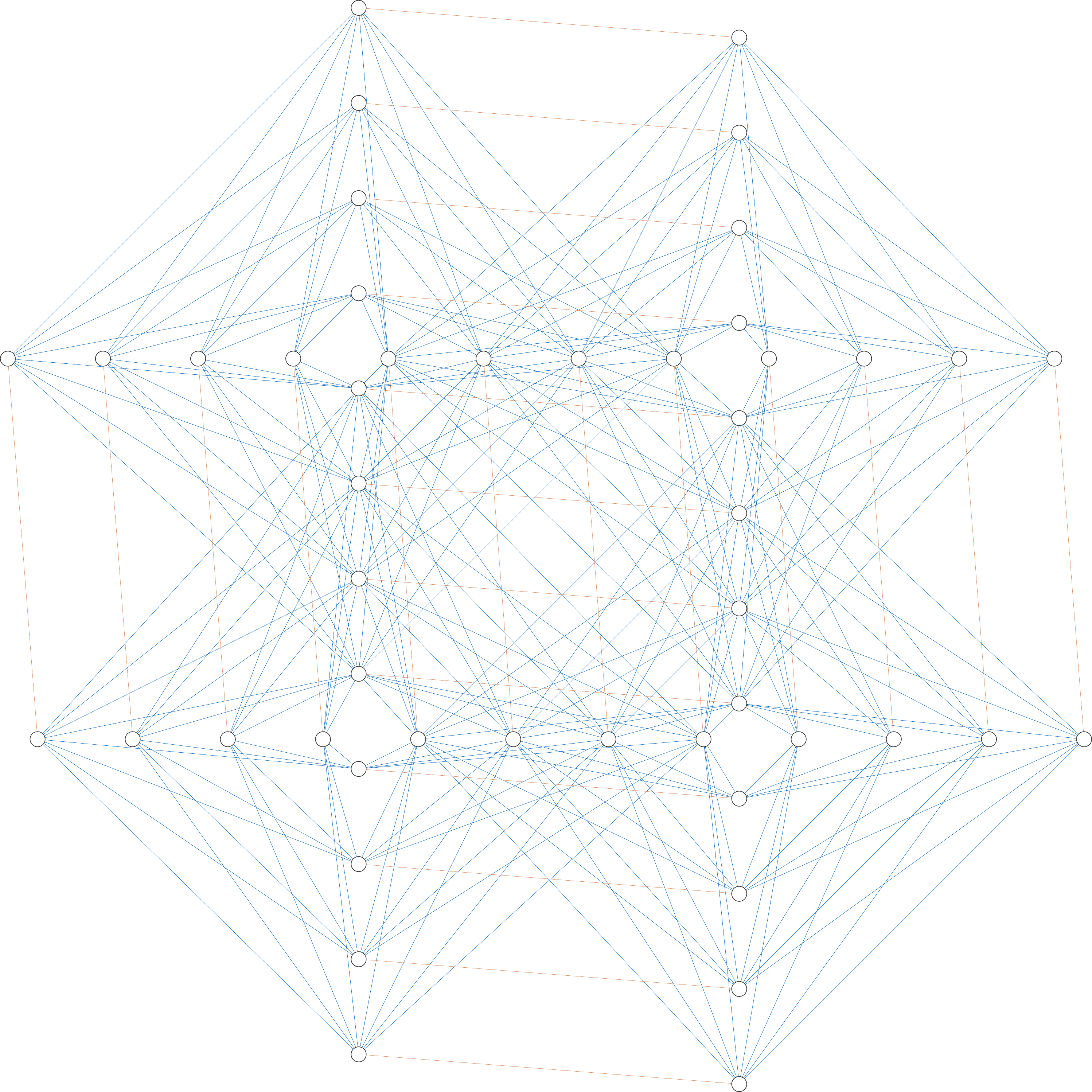}}
\caption{(a) Time-dependent energy scales executed by the D-Wave quantum annealer. The coefficient $A(s)$ (red) multiplies the transverse-field term that enables quantum tunnelling and therefore falls rapidly as the run progresses. The coefficient $B(s)$ (blue) multiplies the problem Hamiltonian and rises smoothly, so that the device moves from a tunnelling-dominated regime at $s=0$ to a problem-dominated regime at $s=1$. (b) Native connectivity map for the Pegasus architecture (D-Wave Advantage). Circles mark individual superconducting qubits; blue lines are programmable couplers that lie within the main module, and orange lines link neighbouring modules. (c) Connectivity map for the newer Zephyr architecture (D-Wave Advantage2), which increases the number of couplers per qubit and introduces longer-range links.}
\label{fig:anneal:overview}
\end{figure*}
The empirical picture is mixed but instructive. D-Wave Advantage2 already delivers an order of magnitude higher ground-state success probability and a smaller time-to-solution exponent compared with its predecessor, signalling rapid hardware progress. Nevertheless, VeloxQ remains the overall leader on the problem sizes accessible today by quantum hardware, underscoring the strength of optimised classical algorithms and the need for denser qubit connectivity or error-mitigation strategies before a categorical quantum advantage can be claimed. Through this work, we provide an open-source practical benchmark suite that spans single-qubit rotations, entangling gates, and PT-symmetric non-Hermitian generators, to supply the community with a transparent yardstick against which future hardware and algorithmic advances can be measured.

While our QUBO encoding builds on the parallel-in-time construction introduced in Ref.~\cite{jalowiecki2020parallel}, the present work addresses a fundamentally different question. Rather than demonstrating feasibility on early-generation hardware, we establish a solver-agnostic, physics-derived benchmarking framework for quantum dynamics and use it to quantitatively assess the current performance gap between modern quantum annealers and state-of-the-art classical heuristics. In contrast to Ref.~\cite{jalowiecki2020parallel}, we consider non-Hermitian and PT-symmetric generators, systematically distinguish native and non-native hardware embeddings, and benchmark two generations of D-Wave processors against GPU-accelerated classical solvers. Unlike Ref.~\cite{veloxq}, which focuses on synthetic or optimization-driven QUBO instances, our benchmarks are derived directly from continuous quantum dynamics, providing a physically motivated and application-relevant testbed for tracking hardware and algorithmic progress.

\section{Methodology}
Before we formalise the benchmark problems and outline the protocol used to generate each instance, it is essential to introduce the computational solvers that underpin all subsequent experiments. Accordingly, the following two subsections are dedicated to an overview of every solver employed throughout this study, establishing a common performance baseline and clarifying the methodological context in which our results should be interpreted.
\subsection{Quantum Annealing}
Quantum annealing begins by initializing the system in the ground state of a simple driver Hamiltonian $H_D$ whose ground state is the uniform superposition $\ket{+}^{\otimes N} = \frac{1}{\sqrt{2^N}} \sum_{x \in \{0,1\}^N} \ket{x}$ with $N$ being the number of qubits.
The goal is to find the ground state of a problem Hamiltonian
\begin{equation}
H_P = \sum_{i=1}^{N} h_i \widehat\sigma_z^{(i)} + \sum_{i<j} J_{ij} \widehat \sigma_z^{(i)} \widehat \sigma_z^{(j)},
\end{equation}
where $\widehat \sigma_z^{(i)}$ is the Pauli matrix for qubit $q_i$, $h_i$ is the qubit bias and $J_{i,j}$ is the coupling strength between qubits. The ground state of $H_P$ encodes the solution to an Ising optimisation problem. During the annealing process, the system evolves under a time-dependent Hamiltonian 
\begin{equation}
    H_\text{Ising}=\frac{A(s)}{2}H_D+\frac{B(s)}{2}H_P,
\end{equation}
governed by QPU dependent anneal schedules $A(s)$ and $B(s)$ (cf. Fig.~\eqref{anneal_schedule}). If the evolution is sufficiently slow, according to the adiabatic theorem \cite{born1928adiabatic}, the system remains in its instantaneous ground state, and the final state gives the solution to the original problem. In addition to standard forward annealing, we employ cyclic annealing as a refinement strategy for selected QUBO instances~\cite{cyclicannealing2024, cyclicannealing2025, cyclicannealing2022}.
The key idea is to repeatedly ``soften'' the optimisation landscape around a candidate solution and then ``re-freeze'' it.
Operationally, the problem Hamiltonian is kept fixed, while the annealing control parameter $s$ is driven through a short back-and-forth schedule:
starting from a classical spin configuration at $s=1$ (problem-dominated regime), the schedule is partially reversed to an intermediate point
$s_{\mathrm{p}}<1$ to temporarily restore the transverse-field contribution, and then returned to $s=1$ where the state is measured.
By repeating this cycle, the solver can leave shallow local minima while still exploiting information contained in the current best configuration~\cite{cyclicannealing2025Poggi, cyclicannealing2025QNN, Schulz2025learningdriven}.

\begin{figure}[t!]
        \includegraphics[width=\linewidth]{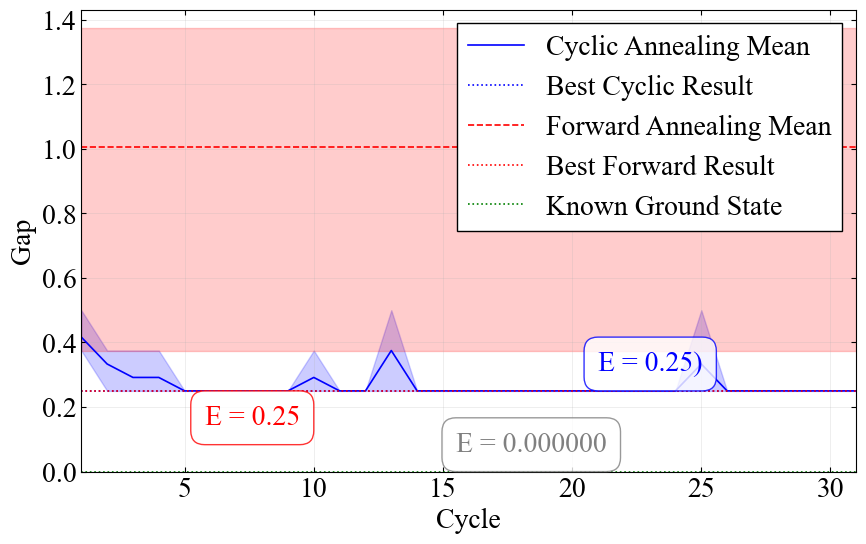}
        
        \caption{Exemplary plateau behavior observed in cyclic annealing for instance $H_3$ with three time points. The blue line indicates the mean over the minimum energy found over each iteration, while the red dashed line shows the mean of the energies found by standard forward annealing. Shadowed box indicates the lower and upper quartile for all samples for forward annealing, and for lowest cycle energy for cyclic annealing. For this plot, we performed three cyclic annealing runs over 30 cycles and 30 forward annealing runs on Advantage2\_system 1.10. The percentage in the box corresponds to the percentage with which lowest energy samples were found by each method.}
        \label{figcyclic}
\end{figure}

A single cycle can be described by a piecewise-linear control schedule $s(t)$,
\begin{equation} s(t)= \begin{cases} 1-(1-s_p)\dfrac{t}{\tau_{\downarrow}}, & 0 \le t < \tau_{\downarrow},\\[3pt] s_p, & \tau_{\downarrow} \le t < \tau_{\downarrow}+\tau_p,\\[3pt] s_p+(1-s_p)\dfrac{t-\tau_{\downarrow}-\tau_p}{\tau_{\uparrow}}, & \begin{aligned}[t] &\tau_{\downarrow}+\tau_p \le t\\ &\le \tau_{\downarrow}+\tau_p+\tau_{\uparrow}, \end{aligned} \end{cases} \tag{3} \end{equation}

where $\tau_{\downarrow}$ and $\tau_{\uparrow}$ set the reverse/forward ramp durations and $\tau_{\mathrm{p}}$ is an optional pause.
The turnaround point $s_{\mathrm{p}}$ controls the locality of the search: larger $s_{\mathrm{p}}$ keeps the dynamics close to the input configuration
(local refinement), whereas smaller $s_{\mathrm{p}}$ induces broader exploration.
We execute $N_{\mathrm{cyc}}$ cycles and keep the lowest objective value observed over the entire cycle sequence.
In practice, cyclic annealing often yields rapid early improvement followed by diminishing returns [cf.\ the plateau behaviour in Fig.~\eqref{figcyclic}],
so the cycle budget is best treated as a tunable trade-off between solution quality and runtime.

D-Wave's superconducting quantum annealers are now applied across a spectrum of physics-driven tasks.  Experiments on a 5000-qubit processor reproduced quantum critical dynamics in 3D spin glasses, extracting critical exponents beyond the reach of classical simulation~\cite{King2023SpinGlass}.  In molecular science, an annealer-based eigensolver has recovered excited-state spectra for small molecules, showing that electronic-structure problems can be recast as Ising models amenable to quantum annealing~\cite{Teplukhin2021Excited}.  Hybrid quantum-classical workflows have also tackled real-time traffic-flow optimisation, demonstrating measurable congestion reduction on realistic road networks~\cite{Neukart2017Traffic, KONIORCZYK2025, domino2024baltimore, smierzchalski2024hybrid, domino2023, ewa2026}. In computer science, an adiabatic implementation of Simon's algorithm has been recently proposed and solved on D-Wave quantum annealers~\cite{robertson2025simonsperiodfindingquantum}. Furthermore, the D-Wave machine has been used as a platform for thermodynamic experiments to validate quantum fluctuation theorems and simulate quantum thermal machines~\cite{Buffoni_2020, Campisi2021, smierzchalski2024efficiency, emery2026}. Progress notwithstanding, widespread deployment is limited by sparse qubit connectivity, which imposes costly minor-embeddings, and by analogue control errors and thermal noise that dilute prospective speed-ups~\cite{Konz2021Embedding, Mehta2025, Vmehta2025, vodeb2024accuracy}.

Due to the limited qubit connectivity in current D-Wave QPUs, problem structures must conform to the device’s \textit{working graph}. The Advantage and Advantage2 QPUs analyzed in this work implement distinct hardware topologies: Pegasus for Advantage and Zephyr for Advantage2 as shown respectively in Figs.~\eqref{pegasus} and \eqref{zephyr}. The Zephyr graph increases native qubit connectivity, with 20 connections per qubit compared to 15 in Pegasus. When a problem instance does not natively fit the hardware graph, it must be embedded via a minor embedding procedure, which may introduce additional complexity. In this study, such embeddings are performed using D-Wave’s heuristic \textit{minorminer} tool when necessary.
\subsection{VeloxQ}
VeloxQ is a fast and efficient, physics-inspired solver based on the theory of dynamical systems, designed for solving QUBO problems, which are crucial for various real-world optimisation tasks~\cite{Lucas_2014}. In a recent work, it has been extensively benchmarked against many state-of-the-art QUBO solvers, D-Wave quantum annealers among them~\cite{veloxq}.  Unlike them (or other physics-inspired optimisation approaches such as gate-based quantum computers), VeloxQ does not rely on technological advancements in hardware to reach its full potential, as it leverages conventional computing technology (such as GPUs). While it is a classical solver, it emulates the probabilistic nature of quantum annealers by producing a low-energy spectrum of solutions. By default, the solver operates with a carefully selected set of parameters that include instance-dependent tuning. However, it also offers flexibility, allowing users to adjust parameters to control the trade-off between solution quality and runtime based on problem knowledge or prior runs. Crucially, it offers stable and repeatable results, which do not suffer from unstable hardware.
VeloxQ distinguishes itself from D-Wave quantum annealers in several significant ways. A key difference lies in their handling of problem topology: D-Wave's quantum processing units (QPUs) have a fixed spin connectivity structure, and necessitate ``minor embedding'' for most real-world problems due to their specific topologies, which introduces substantial overhead. In contrast, VeloxQ natively handles all problem topologies without requiring such embedding. Furthermore, VeloxQ exhibits superior scalability, capable of tackling large-scale optimisation problems with up to $2 \times 10^8$ sparsely connected variables, a feat currently intractable for D-Wave and its hybrid solutions, with estimates suggesting decades before quantum annealers could reach similar capabilities~\cite{veloxq}. 
\subsection{Problem Generation}
\label{sec:problemgeneration}

We briefly describe a method introduced in \cite{jalowiecki2020parallel} that converts the NP-hard problem of solving dynamical systems into finding the ground state of a QUBO formulation, and thus a problem formulation that is suitable to be executed on quantum annealers: Consider an \(L\)--dimensional quantum system evolving under a (possibly non-Hermitian) generator \(H(t) \in \mathbb{C}^{L \times L}\), governed by the first-order differential equation
\begin{equation}
\frac{d}{dt} \ket{\psi(t)} = H(t) \ket{\psi(t)}, \qquad \ket{\psi(t_0)} = \ket{\phi}.
\end{equation}
Upon discretising the time interval \([t_0, t_f]\) into \(N\) steps of size \(\Delta t\), one defines short-time propagators \(U_n = \exp[H(t_n) \Delta t]\). The full evolution is encoded in a so-called history state,
\begin{equation}
\ket{\Psi} = \sum_{n=0}^{N-1} \ket{t_n} \otimes \ket{\psi(t_n)}.
\end{equation}

This state is annihilated by the Hermitian clock operator
\begin{equation}
\mathcal{C} = \sum_{n=0}^{N-2} \left( \ket{t_{n+1}}\bra{t_{n+1}} \otimes I - \ket{t_{n+1}}\bra{t_n} \otimes U_n + \text{h.c.} \right),
\label{clock}
\end{equation}
i.e., \(\mathcal{C} \ket{\Psi} = 0\). To enforce the initial condition \(\ket{\psi(t_0)} = \ket{\phi}\), the following operator is introduced:
\begin{equation}
\mathcal{A} := \mathcal{C} + \ket{t_0}\bra{t_0} \otimes I.
\end{equation}
If the generator \(H(t)\) is Hermitian, then \(\mathcal{A}\) is positive definite, and the solution is given by the minimisation of the quadratic function
\begin{equation}
f(\mathbf{x}) = \frac{1}{2} \bra{\mathbf{x}} \mathcal{A} \ket{\mathbf{x}} - \bra{\mathbf{x}} \phi \rangle.
\end{equation}
The real variables $x_i$ must be converted into binary form so the optimisation can be executed on a quantum annealer (or any other QUBO solver).  Each component is written in fixed-point notation  
\begin{equation}
    x_i = 2^{D}\left(\sum_{\alpha=0}^{R-1}2^{-\alpha}q_i^{\alpha}-1\right),\qquad 
q_i^{\alpha}\in\{0,1\},
\end{equation}
where $R$ bits of precision cover the interval $[-2^{D},2^{D}]$.  Substituting this expression into the quadratic objective $f(\mathbf{x})$ yields a QUBO form,
\begin{equation}
    f(\mathbf{q})=\sum_{i,\alpha} a_i^{\alpha}q_i^{\alpha}
      +\sum_{i,j,\alpha,\beta} b_{ij}^{\alpha\beta}q_i^{\alpha}q_j^{\beta}+f_{0},
      \label{qubo}
\end{equation}
with explicitly computable coefficients,  
\begin{equation}
    \begin{aligned}
b_{ij}^{\alpha\beta} &= A_{ij}\,2^{\,1-\alpha-\beta+2D},\\
a_i^{\alpha} &= 2^{\,1-\alpha+D}\bigl[A_{ii}-2^{D}\bigl(\sum_j A_{ij}-\varphi_i\bigr)\bigr],\\
f_{0}        &= 2^{D}\Bigl(2^{D-1}\!\sum_{ij}A_{ij}+\sum_i\varphi_i\Bigr),
\end{aligned}
\end{equation}
where $A_{ij}$ are elements of the symmetric coefficient matrix $\mathcal{A}$ and quantifies the quadratic coupling between variables $x_i$ and $x_j$. $\varphi_i$ is a component of the linear-term vector and acts as an external bias on the variable $x_i$. The constant shift $f_{0}$ does not influence the minimiser and can be dropped. The edge set mirrors the non-zero pattern of the matrix $A$ derived from the discretised dynamics.  This binary quadratic form is the instance that is finally programmed into the quantum annealer and VeloxQ.
\subsection{Problem Instances}
We examine entangled and non-entangled quantum systems, as well as non-Hermitian Hamiltonians that model quantum dynamics such as $\PT$-symmetric qubits. To perform an accurate performance comparison between classical and quantum solvers, we select dynamical systems that can be precisely modelled as QUBO even with limited precision. Some of these dynamical systems produce QUBO instances that are native to the Pegasus graph, others require an embedding.
\paragraph*{System~1 - Single qubit $Y$-rotation.}
The Hamiltonian is 
\begin{equation}
    H_{1}= \frac{\pi}{2}\,\sigma_{y},
    \label{sys1}
\end{equation}
so that $U(t)=e^{-iH_{1}t}$ enacts a rigid rotation of the Bloch vector about the $y$-axis by an angle $\theta(t)=\tfrac{\pi}{2}t$.  
Eigenvalues $\pm\tfrac{\pi}{2}$ give a fundamental period $T=2$ (in the dimensionless units of the paper), making this an analytically solvable sanity-check instance for the annealer, which is also native to the D-Wave Advantage (Pegasus graph) architecture.
\paragraph*{System~2 - Oblique single qubit drive.}
\begin{equation}
    H_{2}= \frac{\pi}{\sqrt{2}}\bigl(\sigma_{x}+\sigma_{z}\bigr).
    \label{sys2}
\end{equation}
The generator points along the $(1,0,1)$ direction of the Bloch sphere.  
Real eigenvalues $\pm\pi/\sqrt{2}$ yield unitary dynamics that mix computational basis states with a Rabi frequency $\Omega=\pi/\sqrt{2}$, testing the algorithm’s ability to resolve non-orthogonal control axes. Similarly with system 1, Eq.~\eqref{sys1}, this generator is native to the D-Wave Advantage (Pegasus graph) architecture.
\paragraph*{System~3 - Single qubit pair-flip.}
\begin{equation}
    H_3 = \pi \left( \frac{3}{4} \mathbbm{1} - \frac{1}{4} \sigma_y \right).
    \label{sys3}
\end{equation}
This Hamiltonian can be interpreted as a combination of a global phase rotation plus a spin rotation around the y-axis. The off-diagonal terms generate coherent oscillations (Rabi oscillations) between the qubit basis states. The resulting QUBO is not native to the D-Wave Advantage and Advantage2 (Pegasus and Zephyr graphs respectively) architectures.
\paragraph*{System~4 - Native two qubit entanglement.}
\begin{equation}
H_4 = \frac{\pi}{4} \left( \sigma_x \otimes \sigma_x + \sigma_y \otimes \sigma_y \right).
\end{equation}
This Hamiltonian generates maximally entangled Bell states under time evolution. Crucially, its QUBO encoding maps natively to the Pegasus topology of D-Wave Advantage processors, eliminating embedding overhead and providing a direct test of entanglement dynamics on quantum annealers.
\paragraph*{System~5 - Two-qubit pair-flip.}
\begin{equation}
    H_{5}= \frac{\pi}{4} \left( \sigma_x \otimes \sigma_x - \sigma_y \otimes \sigma_y \right).
    \label{sys5}
\end{equation}
Restricted to the $\{\ket{00},\ket{11}\}$ subspace, the spectrum is $\{\pm1\}$ and the dynamics perform Rabi oscillations between these product states.  
At half-period the system passes through the Bell state, providing a clean entanglement-generation benchmark and Pegasus-native problem for the D-Wave Advantage solver.
\paragraph*{System~6 - GHZ-flip on three qubits.}
\begin{equation}
\begin{aligned}
    H_6 &= \frac{\pi}{8} \big( \sigma_x \otimes \sigma_x \otimes \sigma_x 
    - \sigma_x \otimes \sigma_y \otimes \sigma_y \\
    &\quad - \sigma_y \otimes \sigma_x \otimes \sigma_y 
    - \sigma_y \otimes \sigma_y \otimes \sigma_x \big).
\end{aligned}
\label{sys6}
\end{equation}
This is the natural three-qubit analogue of $H_{5}$, Eq.~\eqref{sys5}.  
Evolution swaps $\ket{000}$ and $\ket{111}$ with a period $T=2$ while creating genuine tripartite (GHZ) entanglement that is also Pegasus-native for the D-Wave Advantage device.
\paragraph*{System~7 - $PT$-symmetric qubits.} Besides Hermitian operators, which are usually used as observables in quantum theory, quasi-Hermitian Hamiltonians allow to model unitary evolution as well. We simulate the behavior of a specific type of quasi-Hermitian system, the so-called $\PT$-symmetric Hamiltonians. $\PT$-symmetric systems remain invariant under the combined parity ($\mathcal{P}$) and time reversal ($\mathcal{T}$) operator. Both of these transformations are hermitian and independent of each other, with $[\mathcal{P},\mathcal{T}]=0$ \cite{PhysRevA.94.040101}. Notably, systems with \textit{unbroken} $\PT$ symmetry have a real eigenspectrum and exhibit quasi-Hermitian behavior. A system is unbroken if the eigenstates of $H$ correspond to the eigenstates of the $\PT$ operator. We evaluate the dynamics of the non-Hermitian Hamiltonian

\begin{figure*}
    \centering
    \includegraphics[width=\textwidth]{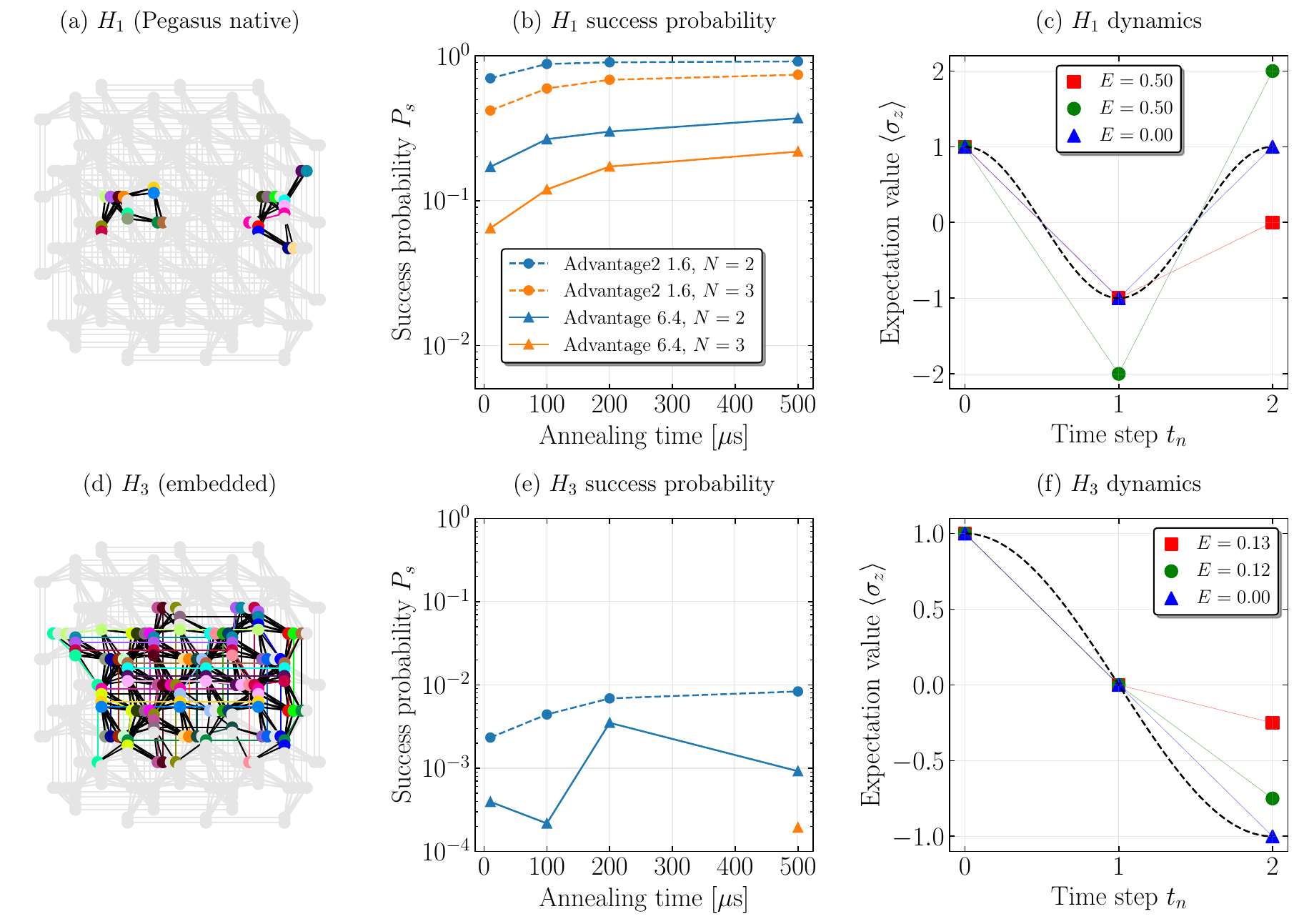}
    \caption{\textbf{Left column}: coloured chains illustrate how the logical qubits of two representative model Hamiltonians are mapped onto the physical Pegasus P16 topology of the D-Wave hardware. Panel (a) shows the native single-qubit problem $H_{1}$, Eq.~\eqref{sys1}, which embeds without chains. Panel (d) shows the non-native single-qubit problem $H_{3}$, Eq.~\eqref{sys3}, where minor embedding produces chains that span several unit cells. \textbf{Middle column}: ground-state success probability obtained on the two quantum annealers-Advantage2 (solid lines) and the earlier Advantage system (dashed lines)-as a function of the annealing time. Circles correspond to a temporal discretisation with two time points, triangles to three. Each marker averages $5000$ annealing runs; if no ground-state sample is observed the marker is omitted. \textbf{Right column}: expectation value $\langle\sigma_{z}\rangle$ reconstructed from the annealer output of the three lowest energy states of the QUBO model of the target Hamiltonian. Blue triangles denote the ground state, green squares and red circles the first and second excited states, respectively. The black dashed curve is the exact unitary evolution computed with QuTiP~\cite{qutip1, qutip2, qutip3, qutip4, qutip5} and serves as a baseline.}
    \label{fig:ta_performance_comparison}
\end{figure*}

\begin{equation}
    H_7 = \omega \left( b \sigma_x + \sin(\alpha) \sigma_y \right), \quad \omega, \alpha, b \in \mathbb{R},
    \label{eq:ptsymmetric}
\end{equation}
with $|\alpha |\!<\!\frac{\pi}{2}$ to ensure that the $\PT$-symmetry of $H$ is unbroken \cite{PhysRevA.91.052113}. $H$ is $\PT$-symmetric with regards to $\mathcal{P}\!=\!\sigma_x=\begin{pmatrix}0&1\\1&0 \end{pmatrix}$ and $\mathcal{T}$ being complex conjugation. With \( b = \sqrt{1 + (\sin \alpha)^2} \), the system's eigenvalues are \( \pm 1 \), allowing for the simulation of its exact time evolution at given time points even with the sparse quantum processing unit graph.
\paragraph*{System~8 - Two qubit system.}
\begin{equation}
    H_8 = \frac{\pi}{4} \left( \mathbbm{1} \otimes \sigma_y - \mathbbm{1} \otimes \sigma_z + \sigma_x \otimes \sigma_x - \sigma_y \otimes \sigma_x \right).
    \label{sys8}
\end{equation}
This Hamiltonian is also entanglement producing, and the resulting QUBO is a connected graph which does not fit natively on the Pegasus and Zephyr graph, i.e. D-Wave Advantage and Advantage2 architectures respectively. 

\section{Performance Evaluation}

\subsection{Sampling Efficiency}
The quality of a quantum annealer is most clearly revealed by the success probability, defined as the fraction of returned bit strings that coincide with the exact ground state of the Ising cost function introduced in Sec.~\ref{sec:problemgeneration}. For every data point we collect \(5000\) annealing runs and repeat the
procedure for four schedule lengths
\(T_{A}\in\{10,100,200,500\}\,\mathrm{\mu s}\).
We distinguish two classes of instances: (i) Native problems that fit directly onto the Pegasus connectivity of the chip, requiring no auxiliary qubits as shown in the upper-left of \autoref{fig:ta_performance_comparison} for system 1, Eq.~\eqref{sys1}. (ii) Embedded problems that violate Pegasus constraints and therefore need minor-embedding chains, for example system 3, Eq.~\eqref{sys3}, as shown in the lower-left of \autoref{fig:ta_performance_comparison}. The middle column of \autoref{fig:ta_performance_comparison} compares the raw success
probabilities obtained on two generations of D-Wave processors and illustrates how the corresponding low-energy samples reconstruct the target wave function. Across the entire data set the newest Advantage2 processor leads the benchmark test. For native problems the median success probability improves by a factor between four and five; for embedded problems the gain grows to one order of magnitude.
Table~\eqref{tab:successratio} summarises these trends, which also confirms observations made for other problem types, such as maximum clique and unweighted maximum cut problems~\cite{Pelofske_2025}.  
\begin{table}[b]
\centering
\begin{tabular}{l|cc}
\toprule
\multirow{2}{*}{$T_A$} & \multicolumn{2}{c}{Success probability ratio (Advantage2 / Advantage)} \\
\cmidrule(l){2-3}
 & native & non-native \\
\midrule
10 & 5.280 & 20.500 \\
100 & 4.778 & 14.000 \\
200 & 4.185 & 4.000 \\
500 & 3.768 & 26.500 \\
\bottomrule
\end{tabular}
\caption{Ratio of success probabilities between Advantage2 and Advantage systems, for Pegasus native and non-native problems. Success probabilities are averaged over 5000 samples and system type (Pegasus native / non-native).}
\label{tab:successratio}
\end{table}
At the shortest schedule (\(T_{A}=10\,\mu s\)) higher success rates---in particular the more than twentyfold gain on non-native problems---point to tighter control of the time-dependent Hamiltonian and reduced calibration error. At intermediate schedules (\(T_{A}=100{-}200\,\mu s\)) the advantage remains, with ratios ranging from about four to fourteen, showing that improvements are not confined to the shortest anneals. At the longest schedule (\(T_{A}=500\,\mu s\)) the advantage grows again, especially for non-native problems. Minor embedding possibly introduces chain breaks and therefore lowers the absolute success probability on \emph{all} processors. Even so, as a result of its greater connectivity, Advantage2 secures ground states roughly four to twenty-five times more often than its direct predecessor, pushing several previously intractable parameter regimes into the range of routine experimentation.
\subsection{Performance Scaling}

We evaluate the samplers' scaling using the time to solution (TTS) metric, which is the computing time required to solve the optimisation problem with a given certainty $P_\text{target}$ or higher. TTS is defined as
\begin{equation}
    \text{TTS}=\frac{\ln(1-P_\text{target})}{\ln(1-P_\text{success})}T_r,
\end{equation}
where $P_\text{success}$ is the probability to find the ground state in a single run and where $T_r$ is the time required for one run. To allow for comparison between classical solver VeloxQ and the D-Wave QPUs, we measure the success probability obtained in one run, which we set to be 1000 samples both on the D-Wave and the VeloxQ. The time used to calculate $\text{TTS}_{99}$ is the time required for one run, averaged over \(5\) runs in the case of D-Wave, and \(20\) runs in the case of VeloxQ and SA.
\begin{figure}[htbp]
        \includegraphics[width=\linewidth]{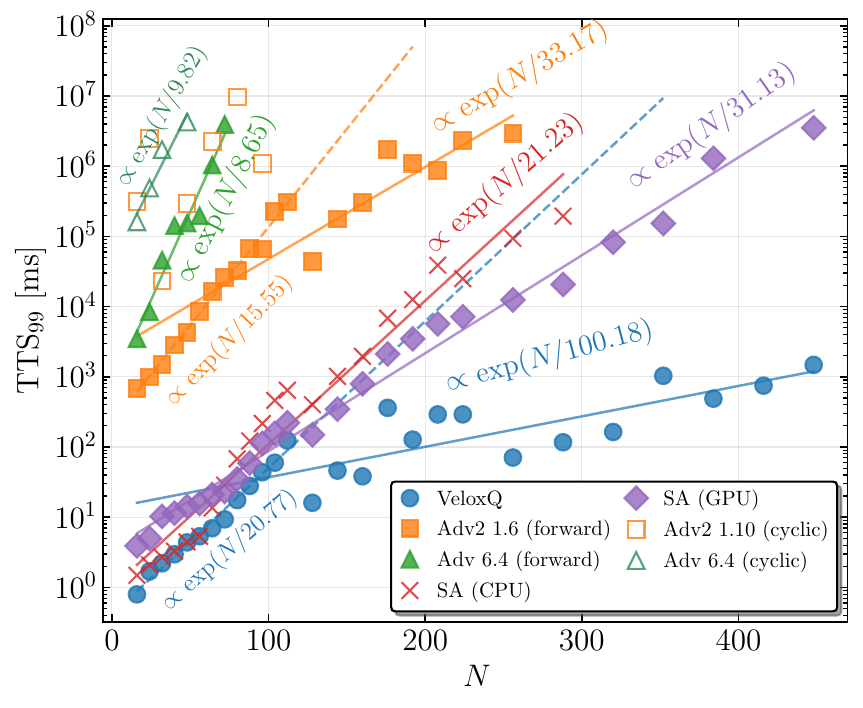}
        \caption{Scaling properties of \(\text{TTS}_{99}\) for Pegasus-native systems \((H_1,H_2,H_5,H_6,H_7)\),~\Cref{sys1,sys2,sys5,sys6,eq:ptsymmetric}, on the Advantage and Advantage2 QPU, as well as the classical solvers: Simulated Annealing~(SA), both CPU and GPU based, and the GPU based VeloxQ. $N$ denotes the number of problem variables. Data points indicate values \(\text{TTS}_{99}\), averaged over system size for all considered systems. Missing data points indicate failure to find the ground state. 5 (20) runs with 1000 samples each were performed on quantum (classical) solvers, for each system at time points $t\in\{2,3,\ldots, 14\}$. }
        \label{fig:tta_overview}
\end{figure}
\autoref{fig:tta_overview} shows scaling of \(\text{TTS}_{99}\) for different instances on quantum (using forward and cyclic annealing) and classical solvers. Since the \(\text{TTS}_{99}\) grows exponentially with the size of the system, we use the fitting function of the form
\begin{equation}
    \text{TTS}_{99} \simeq D \exp\left({\frac{N}{\beta}}\right),
    \label{tts99}
\end{equation}
where $D$ is the function's offset and $N$ is the system size~\cite{Mehta_2022}. The most important fitting parameter is \(\beta\), which can be related to average residual energy density, and subsequently loosely identified with an effective temperature of the generated ensemble of states~\cite{Zhang2025}. In other words, it is a measure of computational complexity associated with a given problem type, and can be used as a quantitative way to compare different solvers, with larger value of \(\beta\) indicating better performance. Of course, one should not expect this quantitiy to be universal across different QUBO problems/spin glass models, with evidence pointing to its dependence on interaction range, and spanning values from up to \(\sim10^3\) in case of Edwards-Anderson model on a cubic lattice~\cite{Zhang2025}, to as low as \(\sim 4.5\) for all-to-all Sherrington-Kirkpatrick model~\cite{Montanaro2020}. In particular, it is not yet fully clear whether quantum solvers (e.g. nonadiabatic quantum annealing) exhibit any inherent advantage over fully classical one.

Figure~\eqref{fig:tta_overview} shows that on the Advantage processor the fitted exponential scale parameter improves from $\beta \simeq 8.65$ (forward) to $\beta \simeq 9.82$ (cyclic),
indicating that cycling can increase the likelihood of reaching the lowest-energy basin, but does not change the overall exponential character of the workload.
This is consistent with the mechanism of cyclic annealing as a guided refinement step: it can help escape nearby suboptimal traps, yet it remains constrained by
the same hardware noise floor and the same energy landscape structure~\cite{cyclicannealing2024}.

Within the overlap region, the cyclic points show substantial scatter and do not exhibit a systematic reduction of $\mathrm{TTS}_{99}$ relative to forward annealing.
A natural interpretation is that the extra control cycles add runtime per attempt, and---once the best reachable energy has been found---additional cycles mainly
accumulate overhead without proportionally increasing the probability of hitting the true ground state.
This interpretation is directly supported by Fig.~\eqref{figcyclic}, where the best-found objective value drops quickly in the first few cycles and then plateaus.

To allow for comparison between quantum and classical solvers we measure the system size as the number of variables before embedding on the QPU's working graph. Notably, the Advantage2 QPU not only has a  \(\text{TTS}_{99}\) that is smaller than the  \(\text{TTS}_{99}\) of its predecessor, but it also exhibits a better scaling exponent. It exhibits two scaling regimes, \(N\lesssim 100\) (\(\beta \approx 15\)) and \(N\gtrsim 100\) (\(\beta \approx 33\)), compared to the uniform Advantage QPU's scaling exponent of $\beta \approx 9$, for the instances that are native to the QPU's working graph (cf. \autoref{fig:tta_overview}). In the latter regimes, its scaling behavior even matches that of an effcient, GPU-based implementation of Simulated Annealing.

In the case of classical solvers, one should again distinguish two regimes: \(N\lesssim 100\) and \(N\gtrsim 100\). In the former, all classical solvers behave similarly in terms of scaling, as well as the individual  \(\text{TTS}_{99}\) values. In the latter, the advantage of GPU over CPU becomes clearly visible, with GPU based SA outperforming the CPU one. Nevertheless, VeloxQ's highly optimized GPU parallelization allows it to take the lead with scaling exponent \(\beta \approx 100\).
For instances that are neither Pegasus nor Zephyr native we are also not able to evaluate the scaling performance, because problem instances quickly become intractable for the QPU (cf. system resolved \(\text{TTS}_{99}\) plots in Supplemental Materials).

Modern quantum annealers are limited by the sparse links that connect their qubits. This limitation is clear in our tests on the single qubit model $H_3$, Eq.~\eqref{sys3}, and the two qubit model $H_8$, Eq.~\eqref{sys8}. The hardware reached the true ground state only twice for $H_3$ and never for $H_8$. The classical algorithm, VeloxQ, faces no such connectivity constraints and solved the same dynamical problems at every time step. Our results reinforce earlier reports showing that VeloxQ outperforms Pegasus- and Zephyr-based quantum processors on problems that do not match their native coupling graphs~\cite{veloxq}. 

\begin{figure}[t!]
    \centering
    \includegraphics[width=\linewidth]{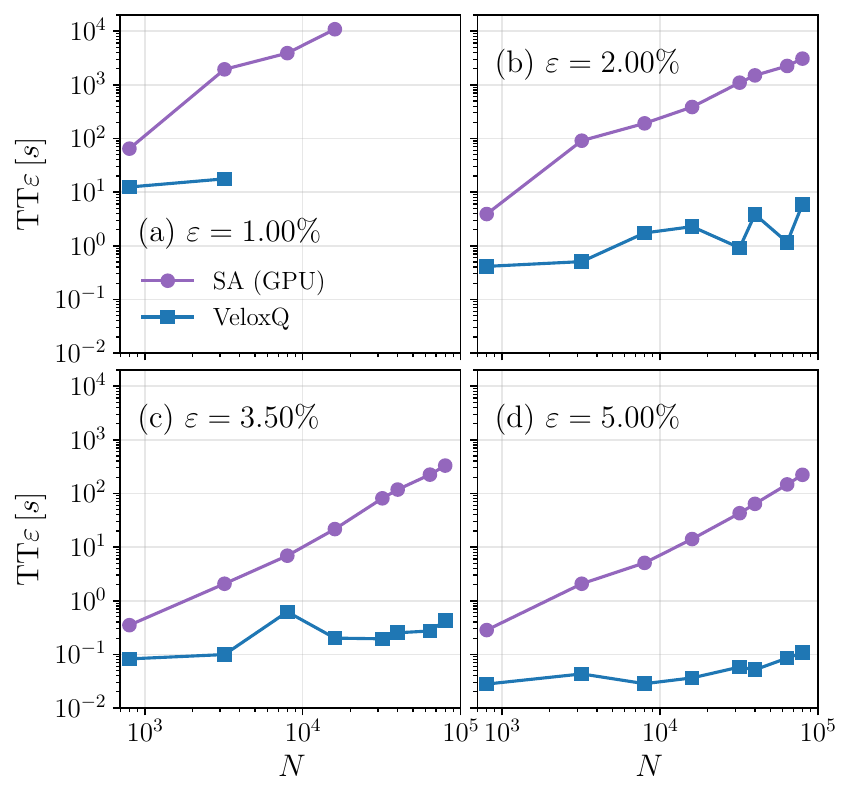}
    \caption{
Scaling of time-to-epsilon $\mathrm{TT}\varepsilon$ for GPU-based classical heuristics, VeloxQ and SA, on large-scale QUBO instances ($N$ up to $10^5$), generated from systems $H_1, H_2, H_6$. Both algorithms display characteristic power-law scaling, where the steepness of the slope is inversely related to the target optimality gap $\varepsilon$. This behavior indicates a crossover toward exponential complexity as the target error vanishes ($\varepsilon\to 0$). For almost all approximate targets shown ($\varepsilon\in\{1.0,2.0,3.5,5.0\}\%$), VeloxQ demonstrates a consistent advantage over Simulated Annealing. The only exception is for $\varepsilon=1.0\%$, where VeloxQ fails to reach the target already for smaller instances than SA.}
    \label{fig:TTe}
\end{figure}

Finally, we extend our benchmarks beyond instances that fit (even with embedding) on current generation annealers. This allows us to establish a classical baseline, within the framework introduced in this work, against which the performance of future quantum devices might be measured. We select three Pegasus-native systems $(H_1, H_2, H_6)$, and construct corresponding QUBO instances with number of discretization timepoints $\{100, 1000, 2000, 5000, 10000\}$ (systems $H_1, H_2$) and $\{100,1000,2000\}$ (system $H_6$). This results in QUBO instances with sizes ranging from $N=800$ to $N=80000$ variables. Since it is infeasible for a heuristic solvers to obtain the true ground state in reasonable time, we relax our performance gauge and use the so-called time-to-epsilon $\mathrm{TT}\varepsilon$~\cite{lidar_advantage, response_to_lidar}, which measures the expected time needed to reach a solution with optimality gap at most $\varepsilon$,
\begin{equation}
  \label{eq:TTe}
  \mathrm{TT}\varepsilon = \frac{\log(1-0.99)}{\log(1-P_{E \leq E_0 + \varepsilon |E_0|})} T_r.
\end{equation}
For the comparison, we consider only the GPU-based VeloxQ solver, and the GPU version of Simulated Annealing. To estimate the success probabiltiy, we performed $100$ independent runs in the former case, and $25$ in the latter. Obtained values of $\mathrm{TT}\varepsilon$ were then optimized over the relevant control parameters of the solvers, them being the number of steps and simultaneously evolved trajectories for VeloxQ, and number of Monte Carlo Sweeps and simultaneously evolved replicas for SA. The results are shown in Figure~\eqref{fig:TTe}. Both solvers exhibit power-law like scaling, with slope increasing with decreasing $\varepsilon$, and eventually diverging to infinity as $\varepsilon\to 0$. This indicates the transition to the exponential scaling of time-to-solution and is consistent with results presented in Figure~\ref{tts99}.
We see that for all considered optimality targets $\varepsilon\in \{1.0, 2.0, 3.5, 5.0\}\%$, VeloxQ produces solutions of desired quality significantly faster (on average), than Simulated Annealing. The only exception is the case of strictest target of $\varepsilon=1\%$, where both solvers cannot reach it for the largest instances, but VeloxQ fails already for smaller systems than SA. Nevertheless, VeloxQ proves itself as a capable solver, especially in the realm of approximate optimization of very large problems.

\section{Conclusion}
We introduced a workflow that recasts continuous quantum dynamics into a sequence of binary optimisation tasks.  Those tasks can be processed without modification on two very different kinds of hardware: superconducting quantum annealers and classical quadratic unconstrained binary optimisation solvers.  The common encoding allowed us to perform a direct, quantitative comparison between the newest D-Wave Advantage2 processor and the GPU-accelerated classical solver VeloxQ.

For the set of dynamical problems tested, D-Wave Advantage2 produced the correct ground-state answer roughly ten times more often than its predecessor, and the increase held across all annealing schedules that we explored.  In practical terms the probability of success rises while the time required to reach a fixed 99 \% confidence level-often used as an application benchmark-grows far more slowly with problem size than on the earlier chip.  These gains mirror the ongoing improvements in qubit coherence, control precision, and on-chip connectivity delivered by the latest fabrication cycle. We additionally tested cyclic annealing as a refinement step for selected instances and found that it typically improves the best objective rapidly in the first few cycles before saturating. 
At the aggregate level, this translates into only modest changes in time-to-solution scaling. 
These observations indicate that cyclic annealing is most useful as a targeted local-improvement heuristic, while the dominant performance gains in our benchmark are driven by the hardware generation upgrade to Advantage2.

At the same time the classical reference solver, VeloxQ, retains the shortest absolute runtime.  Optimised data structures, problem-specific heuristics, and the massive parallelism of modern graphics cards still outweigh the quantum speedups that current devices can offer.  This balance of power illustrates a central theme in contemporary computing: hardware breakthroughs and algorithmic ingenuity progress on separate tracks, and leadership can pass from one to the other multiple times before a mature technology emerges.

The prospects for quantum hardware are encouraging.  Our mapping requires additional qubits only in proportion to the number of simulated time steps, rather than the square or cubic growth familiar from many other encodings. As denser qubit lattices and wider coupler graphs move from prototype to production, medium-scale dynamical simulations covering tens to hundreds of effective degrees of freedom should become accessible within a single annealing run.  Whether the next generation of processors merely matches or overtakes the best classical heuristics will depend on continued advances in both camps. 

Our large-scale benchmarks ($N$ up to $\sim 10^5$) confirm that VeloxQ significantly outperforms Simulated Annealing, exhibiting favorable power-law scaling. This establishes a high-performance classical baseline that helps to indentify the threshold for future quantum advantage.
The open-source benchmark suite released with this work provides a transparent reference against which that progress can be measured.
\section*{Acknowledgements}
The authors acknowledge the \href{https://www.fz-juelich.de/ias/jsc}{J\"ulich Supercomputing Centre} for providing computing time on the D-Wave Advantage™ System JUPSI through the J\"ulich UNified Infrastructure for Quantum computing (JUNIQ). 

P.H. and B.G. acknowledges funding from the National Science Centre (NCN), Poland, under Projects: Sonata Bis 10, No. 2020/38/E/ST3/00269. Z.M. acknowledges funding from the Ministry of Economic Affairs, Labour and Tourism Baden-Württemberg in the frame of the Competence Center Quantum Computing Baden-Württemberg (project ``KQCBW25''). 
\section*{Data Availability}
The source code to reproduce the results and all the data used to generate the plots is available in the accompanying GitHub repository \cite{DWaveDynamics2025}. 
\section*{Image Attribution}
\vspace{0.5em}
\noindent
\href{https://www.flaticon.com/free-icons/gpu-mining}{Gpu mining icons created by Ahmad Roaayala - Flaticon}
\bibliographystyle{apsrev4-2}
\bibliography{bib}

\newpage
\phantom{a}
\newpage

\setcounter{figure}{0}
\setcounter{equation}{0}
\setcounter{page}{1}
\setcounter{section}{0}

\renewcommand{\thetable}{S\arabic{table}}
\renewcommand{\thefigure}{S\arabic{figure}}
\renewcommand{\theequation}{S\arabic{equation}}
\renewcommand{\thepage}{S\arabic{page}}

\renewcommand{\thesection}{S\arabic{section}}

\onecolumngrid

\begin{center}
{\large \bf Supplemental Material:\\
Solving quantum-inspired dynamics on quantum and classical annealers}.\\
\vspace{0.3cm}
Philipp Hanussek$^{1,2}$, Jakub Paw\l{o}wski $^{3,4}$, Zakaria Mzaouali$^{5,6}$, Bart\l{o}miej Gardas$^1$\\
$^1$ {\it Institute of Theoretical and Applied Informatics, Polish Academy of Sciences, Ba\l{t}ycka 5, Gliwice, 44-100, Poland.}\\
$^2$ {\it Sorbonne Université, Faculté des Sciences et Ingénierie, 75005 Paris, France.}\\
$^3$ {\it Institute of Theoretical Physics, Faculty of Fundamental Problems of Technology, \\ Wroc\l{a}w University of Science and Technology, 
50-370 Wroc\l{a}w, Poland.}\\
$^4$ {\it Quantumz.io Sp. z o.o., Pu\l{a}wska 12/3, 02-566, Warsaw, Poland.}\\
$^5$ {\it Institut für Theoretische Physik, Universität Tübingen, Auf der Morgenstelle 14, 72076 Tübingen, Germany.}\\
$^6$ {\it J\"ulich Supercomputing Centre, Institute for Advanced Simulation, Forschungszentrum J\"ulich, Wilhelm-Johnen-Stra{\ss}e, J\"ulich, 52428, Germany.}
\end{center}

\section{System-resolved time-to-solution results}
\begin{figure}[htpb]
  \centering
  \vspace{-0.3cm}
  \begin{subfigure}[t]{0.3\textwidth}
    \caption{System $H_1$}
    \includegraphics[width=\linewidth]{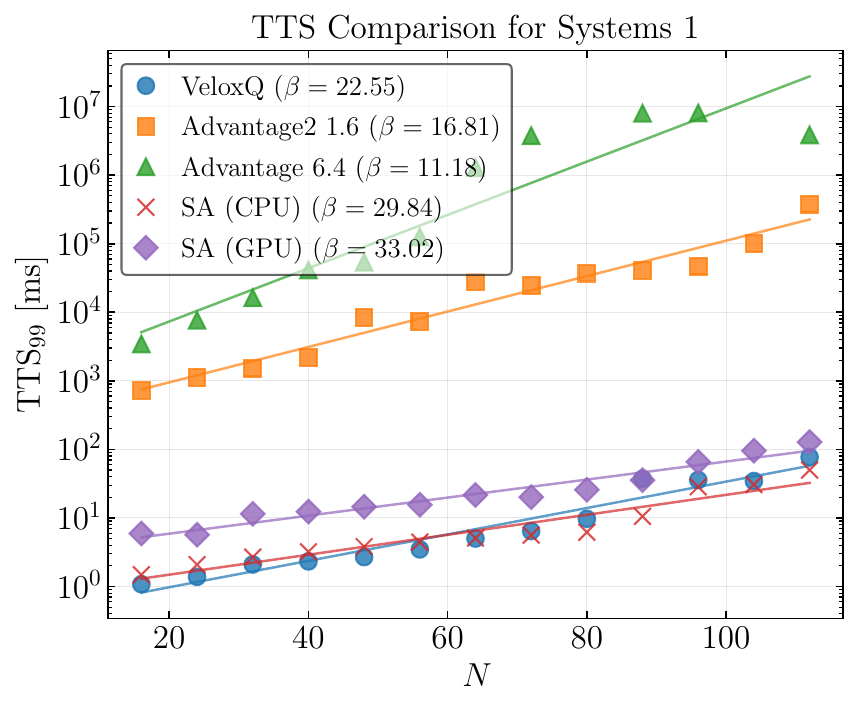}
  \end{subfigure}
  \begin{subfigure}[t]{0.3\textwidth}
    \caption{System $H_2$}
    \includegraphics[width=\linewidth]{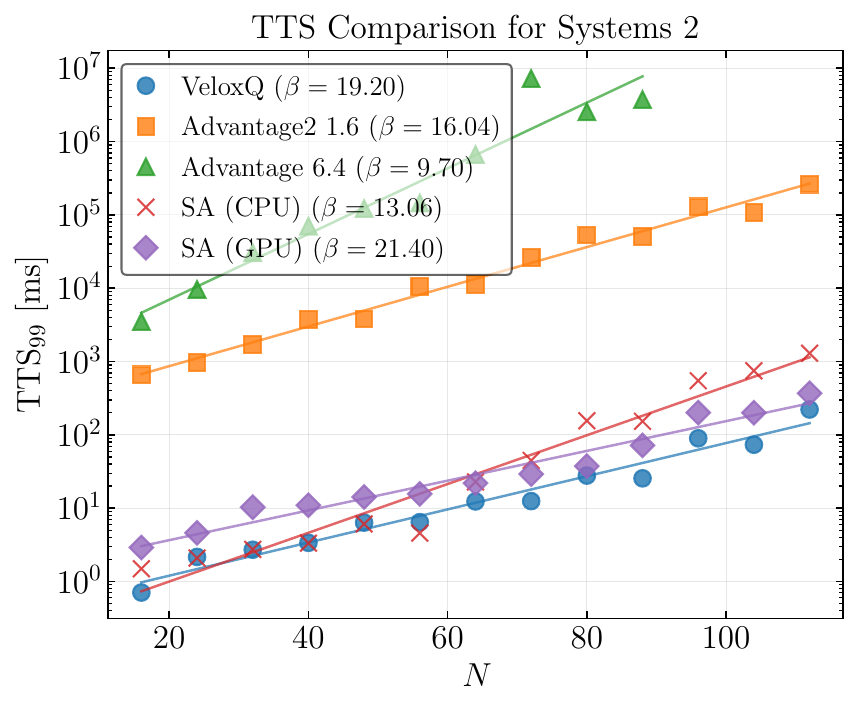}
  \end{subfigure}
  \begin{subfigure}[t]{0.3\textwidth}
    \caption{System $H_3$}
    \includegraphics[width=\linewidth]{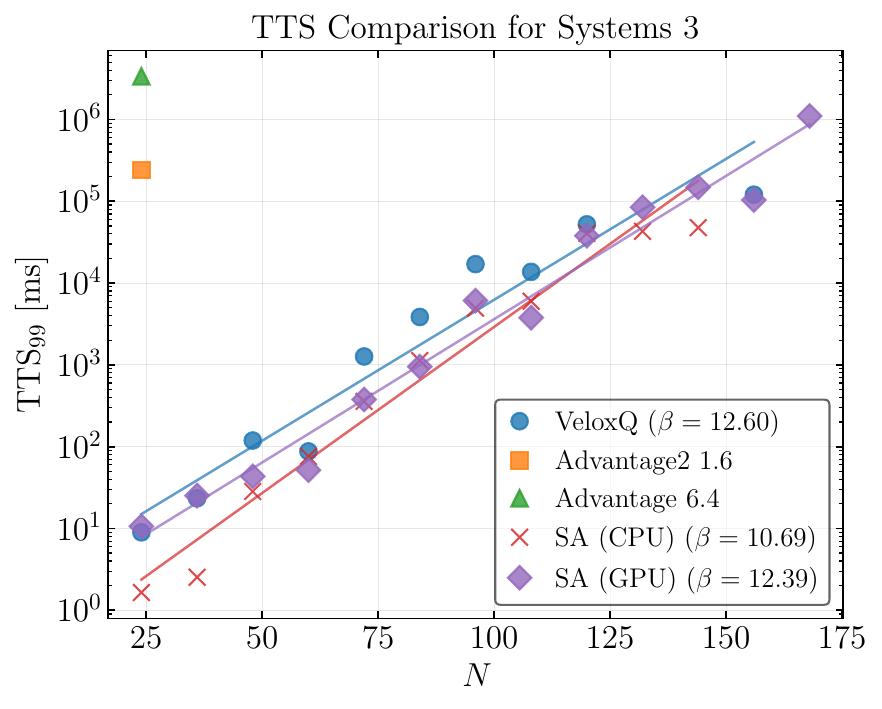}
  \end{subfigure}
\vspace{-0.3cm}
  \begin{subfigure}[t]{0.3\textwidth}
    \caption{System $H_4$}
    \includegraphics[width=\linewidth]{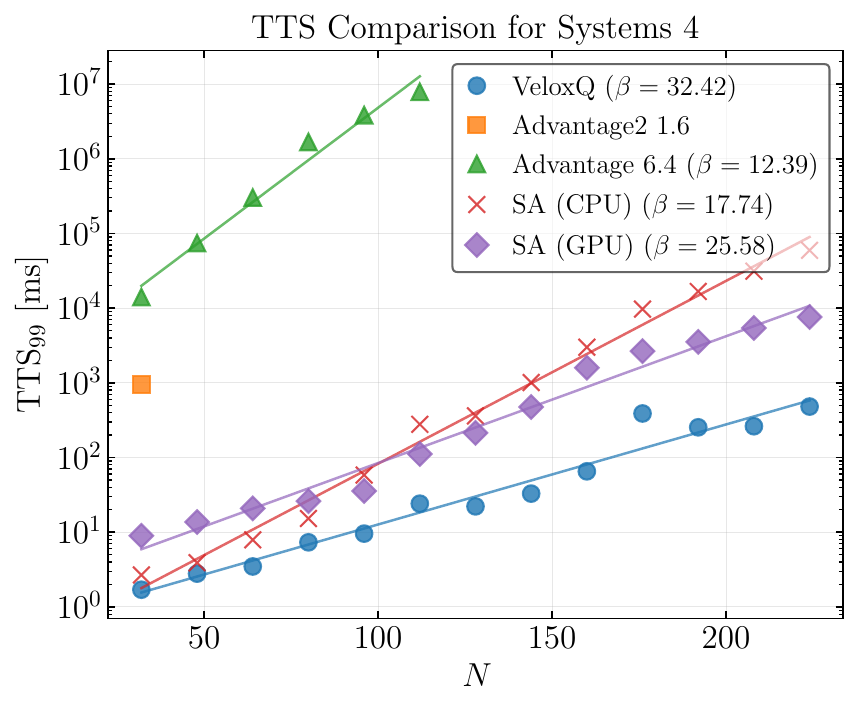}
  \end{subfigure}
  \begin{subfigure}[t]{0.3\textwidth}
    \caption{System $H_5$}
    \includegraphics[width=\linewidth]{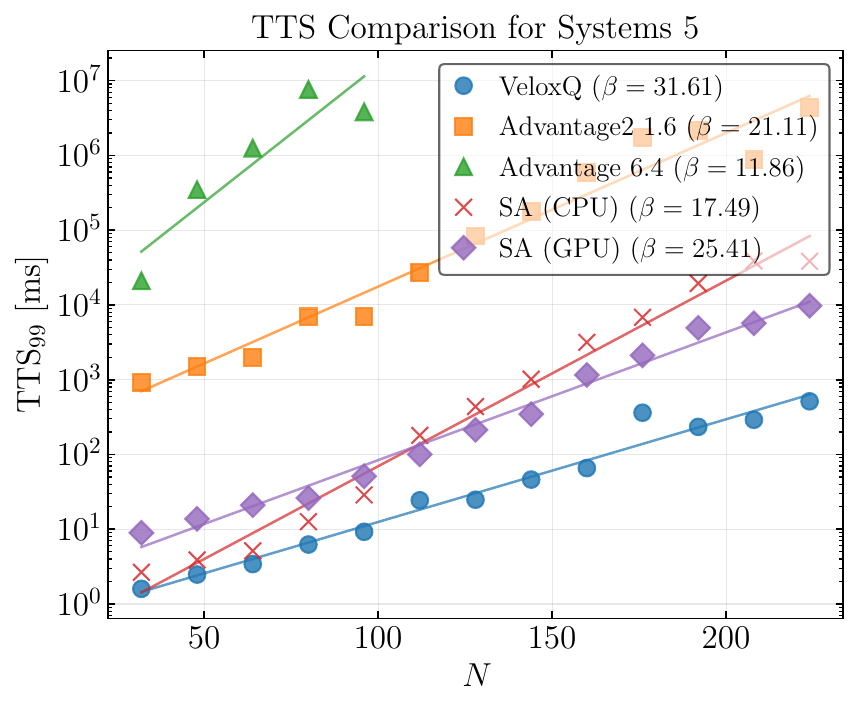}
  \end{subfigure}
  \begin{subfigure}[t]{0.3\textwidth}
    \caption{System $H_6$}
    \includegraphics[width=\linewidth]{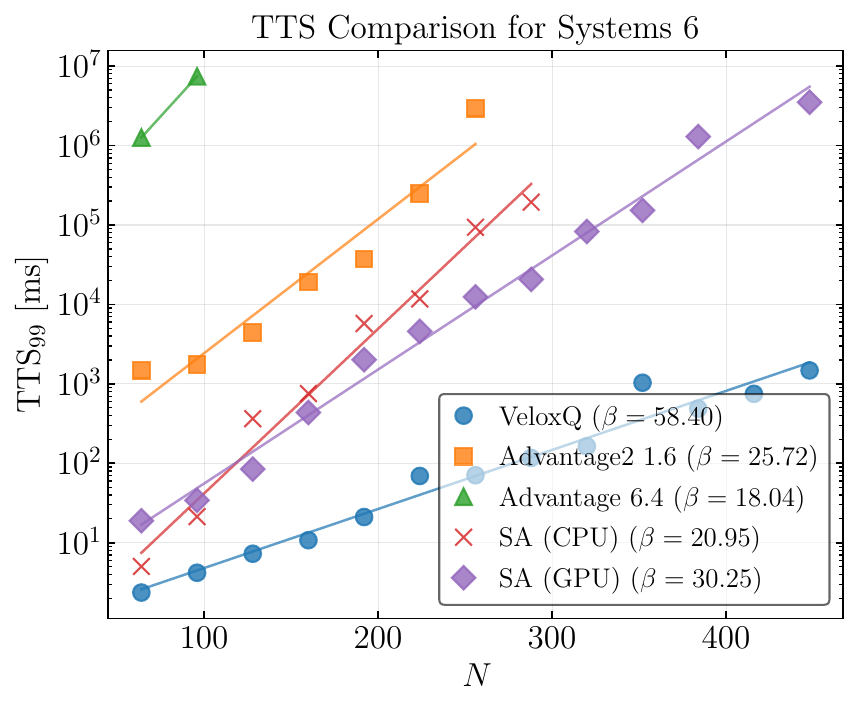}
  \end{subfigure}
\vspace{-0.3cm}
  \begin{subfigure}[t]{0.3\textwidth}
    \caption{System $H_7$}
    \includegraphics[width=\linewidth]{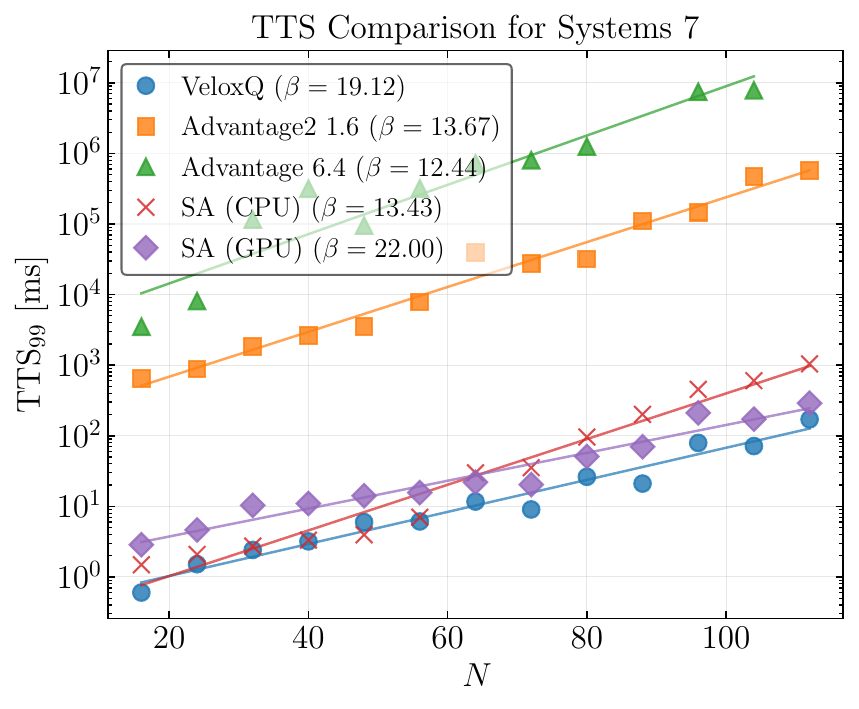}
  \end{subfigure}
  \begin{subfigure}[t]{0.3\textwidth}
    \caption{System $H_8$}
    \includegraphics[width=\linewidth]{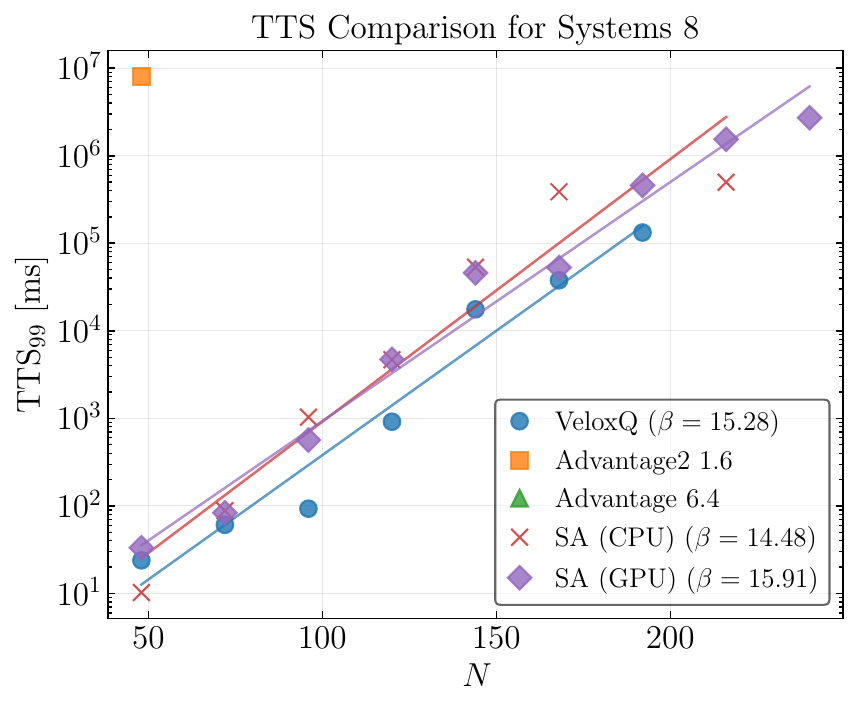}
  \end{subfigure}

  \caption{Plots of system-resolved \(\text{TTS}_{99} \) as a function of system size, demonstrating clear exponential scaling \(\propto \exp(N/\beta)\), with \(\beta\) exponent indicated in the legend. Missing data points indicate zero success probability, which is the case e.g. for quantum solver in systems \(H_3\) and \(H_8\). In the \(N \gtrsim 100\) regime the advantage of GPU solvers becomes clearly visible. Nevertheless, the new Advantage2 1.6 QPU demonstrates the developing potential of quantum annealers.}
  \label{fig:8plots}
\end{figure}

\end{document}